\documentclass[conference]{IEEEtran}
\IEEEoverridecommandlockouts
\usepackage[utf8]{inputenc}
\usepackage{bm}
\usepackage{amsmath}
\usepackage{amsthm}
\usepackage{amsfonts}
\usepackage{mathrsfs}
\usepackage{pifont}
\usepackage{amssymb}
\usepackage{verbatim}
\usepackage{upgreek}
\usepackage{color}
\usepackage{graphicx}
\usepackage{caption}
\usepackage{subfigure}
\usepackage{algorithmic}
\usepackage{algorithm}
\usepackage{epsfig}
\usepackage{wrapfig}
\usepackage{cite}
\usepackage{setspace}
\usepackage{multirow}
\usepackage{fancyhdr}
\usepackage{graphicx} %use graph format
\usepackage{epstopdf}
\usepackage{makecell}
\usepackage{stfloats}
\usepackage{booktabs}
\usepackage{array}
\usepackage{balance}

\def\BibTeX{{\rm B\kern-.05em{\sc i\kern-.025em b}\kern-.08em
T\kern-.1667em\lower.7ex\hbox{E}\kern-.125emX}}
\begin{document}
\title{Weak Target Detection with Multi-bit\\Quantization in Colocated MIMO Radar
}
\author{
%\IEEEauthorblockN{Hang~Xiao, Shixing~Yang, and Wei~Yi}
%\IEEEauthorblockA{\textit{School of Information and Communication Engineering} \\
%\textit{University of Electronic Science and Technology of China}\\
%Chengdu, China\\
%Email: hangxiaoxh@163.com}\\

\IEEEauthorblockN{Hang~Xiao, Shixing~Yang, and Wei~Yi}
\IEEEauthorblockA{\textit{University of Electronic Science and Technology of China, Chengdu, China} \\
Email: hangxiaoxh@163.com, yangshixing@std.uestc.edu.cn, kussoyi@gmail.com}\\

}
\maketitle
\begin{abstract}
We consider the weak target detection problem with unknown parameter in colocated multiple-input multiple-output (MIMO) radar. To cope with the sheer amount of data for large-size systems, a multi-bit quantizer is utilized in the sampling process.  As a low-complexity alternative to classic generalized likelihood ratio test (GLRT) for quantized data, we propose the multi-bit detector on Rao test with a closed-form test statistic, whose theoretical asymptotic distribution is provided to generalize the actual detection performance. Additionally, we refine the design of quantizer by optimized quantization thresholds, which are obtained resorting to the popular particle swarm optimization algorithmthe (PSOA). The simulation is conducted to demonstrate the performance variations of detectors based on unquantized and quantized data. The numerical results corroborate our theoretical analyses and show that the performance with 3-bit quantization approaches the case without quantization.
\end{abstract}
\section{Introduction}
Multiple-input multiple-output (MIMO) radar, which consists of multiple transmit antennas and multiple receiver antennas with waveform diversity, has been heavily investigated for years \cite{MIMO1,MIMO2,MIMO3,MIMO4,MIMO5}. In particular, through the combination of digital array signal processing, it can dramatically improve the performance of target detection, especially in low signal-to-noise ratio (SNR). However, with the expansion of array scale, it requires powerful computation capability to perform real-time detection for the mass receive data. That increases the difficulty of hardware implementation as well as the load of data transmission. To address those challenges in practice, the received data may be quantized before further transmission and processing. %按照参考文献的顺序再排一下

In the field of wireless sensor networks, one-bit quantization, as the simplest and humblest design of quantizer, has attracted significant attention due to its properties of satisfying the stringent bandwidth and energy constraints \cite{ss1,ss2,ss3}. Obviously,  as a result of the damage to signal integritythe, the quantization will lead a performance degradation in some applications, such as target detection and localization. A number of studies have succeeded in measuring the loss and finding ways to compensate for it. Consider the additive white Gaussian noise (AWGN) environment, an important conclusion is that the estimated variance based on one-bit quantized data is as small as $\pi/2$ times that of the clairvoyant sample mean estimator \cite{1bg1}. By noting the conclusion only holds if the quantization threshold has been well-designed, a comprehensive analysis is introduced in \cite{1bg2} and a lot of applications for estimation are derived \cite{1bg3,1bg4}.

Moreover, in the case of detection, a one-bit detector on the generalized likelihood ratio test (GLRT) is proposed in \cite{1bj1}, which is conducted by replacing the unknown parameter with the maximum likelihood estimation (MLE). Then, the Rao test is adopted as the computationally simpler alternative under specified conditions \cite{1bj3}, which indicates that the $\pi/2$ criterion with regard to the performance loss caused by one-bit quantization still approximately holds \cite{1bj2}. Furthermore, considering the substantial loss of information under changeable conditions, multi-bit quantization is studied by trading off performance and complexity \cite{m1,m2,m3}.

It is important to note that the above methods proposed in the real domain are not completely applicable to radar. But recently, by quantizing the real and imaginary parts of the complex signal respectively, increasing efforts have been devoted to studying the performance of radar systems based on quantized data. In \cite{radar1}, a novel approach is proposed for target parameter estimation in cases where one-bit analog-to-digital-converters (ADCs), also known as signal comparators with time-varying thresholds. Then under the Neyman-Pearson criterion, the one-bit likelihood ratio test (LRT) detector is proposed with prior knowledge of reflectivity parameter assumed to be known \cite{radar2}. The one-bit detector on Rao test is also derived as the improvement, which takes the reflectivity parameter as an unknown deterministic value \cite{radar3}.

Nevertheless, to the best of our knowledge, the case of multi-bit quantization for radar detection has not been well carried out. Also, the optimization of quantization thresholds is a valuable point to explore, which has a direct impact on performance with quantization methods. Thus, motivated by \cite{m2}, we extend the one-bit quantization mentioned in \cite{radar3} to the case of multi-bit for radar and explicate the setting of quantization thresholds to refine the design of quantizer.

Specifically, this paper considers a detection problem of weak (in low SNR) target with unknown parameter in colocated MIMO radar. We propose a multi-bit detector on Rao test, which follows a closed form test statistic without computing the MLE. Additionally, the theoretical asymptotic performance is provided as the instruction of optimizing quantizer. By maximizing the corresponding parameter of approximate distribution, the optimized quantization thresholds are obtained resorting to the particle swarm optimization algorithm (PSOA). Simulation results corroborate our theoretical analyses, and demonstrate that the performances exposed by 2-bit and 3-bit detectors are between those of 1-bit and $\infty$-bit (without quantization) detectors. The performance improves with the increase of quantization bits. Especially, the case of 3-bit detector approaches that of $\infty$-bit, which proves the validity of multi-bit quantization.

Notation: We use bold lowercase letters for vectors and bold uppercase letter for matrices. $(\cdot)^\top$ and $(\cdot)^H$ denote the transpose and the conjugate transpose of the vector or matrix argument, respectively. $(\cdot)^{-1}$ denotes the inversion of matrix argument or function argument. $\Re(\cdot)$ and $\Im(\cdot)$ are the real and imaginary parts of a complex vector or scalar, respectively. $\mathbb{E}(\cdot)$, $\mathbb{R}(\cdot)$ and $\mathbb{C}(\cdot)$ denote the expectation, real field and complex field, respectively. $j$ is the imaginary unit. $\left \|\cdot\right \|$ denotes the norm of a vector. $\textbf{I}$ is the identity matrix.
\section{Signal Model}
We consider a colocated MIMO radar system with $N_t$ transmit and $N_r$ receive antennas placed as uniform linear arrays (ULAs). Assume the target appears in the far field of the antenna arrays with $\phi$ denoting the location parameter, the transmit-receive channel matrix may be defined as
\begin{equation}
{\bf{A}}\left( \phi \right)\!=\!{{\bf{a}}_r}\left( \phi \right){\bf{a}}_t^\top\!\left( \phi \right)\!=\! \!
\begin{bmatrix}
\begin{smallmatrix}
1&e^{A_{0,1}}&\cdots&e^{A_{0,N_t\!-\!1}}\\
e^{A_{1,0}}&e^{A_{1,1}}&\cdots&e^{A_{1,N_t\!-\!1}}\\
\vdots&\vdots&\ddots&\vdots\\
e^{A\!_{N_r\!-\!1\!,1}}&e^{A\!_{N_r\!-\!1\!,1}}&\cdots&e^{A\!_{N_r\!-\!1\!,N_t\!-\!1}}
\end{smallmatrix}
\end{bmatrix}\!,
\end{equation}
where $A_{i_r,i_t} = \left[-j2\pi\left({i_r + i_t}\right)d\sin \phi \right]/\lambda$, with the wavelength $\lambda$ and the antenna spacing $d$. ${\bf{a}}_t\left( \phi \right) \in \mathbb{C}^{{N_t} \times 1}$ and ${\bf{a}}_r\left( \phi \right) \in \mathbb{C}^{{N_r} \times 1}$ stand for the transmitted and received array response vectors respectively \cite{MIMO4,MIMO5}.

Since the multi-bit quantizer works in the sampling process, the digital-to-analog converters (DACs) adopted at the transmit antennas are assumed to be infinite-resolution \cite{radar2}. Let the ${\bf{S}}\in \mathbb{C}^{{N_t} \times L}$ ($L$ represents the time dimension) denotes the discrete-time signal transmitted by all transmit antennas synthetically. The received signal at the input of the ADCs can be written as
\begin{equation}
{\bf{X}} = \beta {\bf{A}}\left( \phi \right){\bf{S}} + {\bf{W}},
\end{equation}
where $\beta$ is an unknown complex scalar which stands for the reflection coefficient of target, and ${\bf{W}}\in \mathbb{C}^{{N_t} \times L}$ denotes the additive white Gaussian noise with zero mean and covariance matrix $\sigma ^2\textbf{I}$ \cite{radar3}.

By quantizing the real and imaginary parts of the received signal respectively \cite{radar1, radar2, radar3}, the sampled signal after applying $q$-bit quantizer ${U_q}\left(\cdot\right)$ can be expressed as
\begin{equation}
{\bf{Y}} = {U_q}\left( {\bf{X}} \right) = {U_q}\left({\Re \left( {\bf{X}} \right)} \right) + j {U_q}\left( {\Im \left( {\bf{X}} \right)} \right).
\end{equation}
\quad Through comparing element $x$ of the signal $\bf{X}$ with a set of strictly monotone increasing thresholds $\left\{\tau_k,k = 0,1,...,{2^q} \right\}$ (being $\tau_0=-\infty$ and $\tau_{2^q}=+\infty$) independently, the corresponding output $y$ of the $q$-bit quantizer is encoded as a binary code, which is given by 
\begin{equation}
y = {U_q}\left( { x} \right) \buildrel \Delta \over = \left\{ {\begin{array}{*{20}{c}}
\begin{array}{l}
{b_1}\\
{b_2}\\
\vdots \\
{b_{{2^q}}}
\end{array}&\begin{array}{l}
- \infty < x < {\tau _1}\\
{\tau _1} < x < {\tau _2}\\
\vdots \\
{\tau _{{2^q} - 1}} < x < + \infty 
\end{array}
\end{array}} \right.,
\end{equation}
where $\left\{{{b_k},k = 1,2,...,{2^q}}\right\}$ are the binary codewords with code length $q$. For example, given $q$ = 2, we have ${b_1}$ = ‘00’, ${b_2}$ = ‘01’, ${b_3}$ = ‘10’ and ${b_4}$ = ‘11’ \cite{m3}. Then the binary hypothesis testing problem is formulated as
\begin{equation}
\left\{ \begin{array}{l}
{H_0}:{\bf{Y}} = {U_q}\left( {\bf{X}} \right) = {U_q}\left( {\bf{W}} \right)\\
{H_1}:{\bf{Y}} = {U_q}\left( {\bf{X}} \right) = {U_q}\left( {\beta {\bf{A}}\left( \phi \right){\bf{S}} + {\bf{W}}} \right)
\end{array} \right..
\end{equation}
\quad After centralized processing of the quantized signal, the system would make a final decision about the absence or presence of target. The detector derivation and quantization threshold selection, which both have a direct impact on the detection performance, will be elaborated in the following.
\section{Multi-bit Detector}
A common approach for binary hypothesis testing problem is given by GLRT, which replaces the unknown parameter with the MLE \cite{ss2,1bj1,m1}. To match the proposed quantization method, we define vector ${\boldsymbol{\upbeta}}{\rm{ = }}{\left[ {{\beta _R},{\beta _I}} \right]^{\top}}$ as an equivalent expression of $\beta = {\beta _R} + {\beta _I}j$. Then corresponding test statistic is constructed as :
\begin{equation}
\Lambda _G=\ln{\frac{P\left({\left. {\bf{Y}} \right|\!{H_1};{\boldsymbol{\upbeta}} }\right)}{P\left( {\left. {\bf{Y}} \right|\!{H_0}} \right)}}\Bigg|_{\boldsymbol{\upbeta}=\hat{\boldsymbol{\upbeta}}}\mathop{\mathbin{\lower.3ex\hbox{$\buildrel>\over
{\smash{\scriptstyle<}\vphantom{_x}}$}}}\limits_{{H_0}}^{{H_1}}\eta,
\end{equation}
where ${P\left({\left.{\bf{Y}}\right|\!{H_1};{\boldsymbol\beta} } \right)}$ and $P\left( {\left. {\bf{Y}} \right|\!{H_0}} \right)$ are the probability mass functions (PMF) of quantized data $\bf{Y}$ under $H_0$ and $H_1$ hypotheses respectively, the threshold $\eta$ is determined by a given false-alarm probability, and the MLE of $\boldsymbol\upbeta$ is
\begin{equation}
\hat{\boldsymbol\upbeta} = \arg \mathop {\max }\limits_{\boldsymbol\upbeta} \ln P\left( {{\bf{Y}};{\boldsymbol\upbeta},{H_1}} \right).
\end{equation}
\quad However, the above optimization problem does not allow a closed-form analytical solution usually and numerical solution methods increase the computational complexity of its implementation. Therefore, as a simpler (without computing $\hat{\boldsymbol{\upbeta}}$) alternative to the GLRT, we will refer to the employed Rao test \cite{ss2,1bj2,m2}, whose test statistic is constructed as:
\begin{equation}\label{1}
{\Lambda _R} \!=\! \!{\left[ \!{{{\left( \!{\frac{{\partial\! \ln \!P\!\left( {\left. {\bf{Y}} \right|\!{H_1};\!{\boldsymbol\upbeta} }\! \right)}}{\partial{\boldsymbol\upbeta} }} \right)}^{\!\!\top}}\!\!{\bf{F}}{{\bf{I}}^{ - \!1}}\!\!\left( {{{\boldsymbol\upbeta}}} \right)\!\left( {\frac{{\partial \!\ln\! P\!\left(\! {\left. {\bf{Y}} \right|\!{H_1};\!{\boldsymbol\upbeta} } \right)}}{\partial{\boldsymbol\upbeta} }} \!\right)}\! \right]\!\!\Bigg|_{{\boldsymbol\upbeta} = {{\boldsymbol\upbeta} _0}}},
\end{equation}
where ${{\boldsymbol\upbeta}_0}= {\left[ {0,0} \right]^{\top}}$ is the true value of ${\boldsymbol\upbeta}$ under $H_0$, and ${\bf{FI}}\left({\boldsymbol{\upbeta}}\right)\in {\mathbb{R}}{^{2 \times 2}}$ is the Fisher information matrix, i.e.,
\begin{equation} 
{\bf{FI}}\left( {{{{\boldsymbol\upbeta}}_0}} \right) = \mathbb{E}{\left[{{\rm{}}{{\left( {\frac{{\partial \ln P\left( {\left. {\bf{Y}} \right|{H_1};{{\boldsymbol\upbeta}}} \right)}}{{\partial {{{\boldsymbol\upbeta}}}}}} \right)}^2}}\right] \Bigg|_{{{\boldsymbol\upbeta}} = {{{\boldsymbol\upbeta}}_0}}},
\end{equation}
\quad In order to obtain ${\Lambda _R}$ explicitly, we define ${\bf{Z}} = {\bf{A}}\left( \phi \right){\bf{S}} = {\bf{G}} + j{\bf{H}}$. Then we have a more intuitive expression of $\beta \bf{Z}$ by considering the matrix as ${N_r}L$ elements \cite{radar2 ,radar3}, that is
\begin{equation}
\begin{aligned} 
\beta {z_n} = {\beta _R}&{g_n} - {\beta _I}{h_n} + j({{\beta _R}{h_n} + {\beta _I}{g_n}})\\&n = 1,2,...,{N_r}L
\end{aligned}.
\end{equation}

Based on above expression, we define $y_n$ as the $n$-th element of the quantized signal $\bf{Y}$. According to the noise distribution and the quantization method, the PMF of $\Re(y_n)$ is given by
\begin{equation} 
\begin{split} 
&P\left( {\Re \left( {{y_n}} \right) = b_i;{\boldsymbol\upbeta}} \right) = P\left( {{\tau _{i - \!1}} \le \Re \left( {{y_n}} \right) < {\tau _i};{\boldsymbol\upbeta}} \right)\\
&=\!Q\!\!\left(\!\frac{\tau_{i-\!1}\!-\!({\beta_R}{g_n}\!-\!{\beta_I}{h_n})}{\sigma_n/\sqrt2}\!\right)\!\!-\!Q\!\!\left(\!\frac{\tau_i\!-\!({\beta_R}{g_n}\!-\!{\beta_I}{h_n})}{\sigma_n/\sqrt2}\!\right)\\
&=F_{n,i}({\beta_R}{g_n}\!-\!{\beta_I}{h_n})
\end{split},
\end{equation}
where $Q(\cdot)$ is the complementary cumulative distribution function of the standard normal distribution \cite{ss2}. Similarly, the PMF of $\Im({y_n})$ is given by
\begin{equation} 
P\left( {\Im \left( {{y_n}} \right) = b_i;{{\boldsymbol\upbeta}}} \right) = {F_{n,i}}\left( {{\beta _R}{h_n} + {\beta _I}{g_n}} \right).
\end{equation}

Exploiting the independence of sampling sequence and receiving channels, we expand $\ln P\left( {\left. {\bf{Y}} \right|{H_1};{\boldsymbol{\upbeta }}} \right)$ as
\begin{equation} 
\begin{split}
\ln P\!\left( {\left. {\bf{Y}} \right|{H_1};{\boldsymbol{\upbeta }}} \right)\!=\!&\sum_{n=1}^{N_r\!\times\! L}\!\left[\ln\sum_{i=1}^{2^q}I_i\left(\Re(y_n)\right)F_{n,i}({\beta_R}{g_n}\!-\!{\beta_I}{h_n})\right.\\
&+\left.\ln\sum_{i=1}^{2^q}I_i\left(\Im(y_n)\right)F_{n,i}({\beta_R}{h_n}\!+\!{\beta_I}{g_n})\right],
\end{split}
\end{equation}
where $I_i\left(y\right)=1$ if $y=b_i$ and 0 otherwise. Also, the $P\left(\! {\left. {\bf{Y}} \right|{H_0}} \right)$ could be easily obtained as $\boldsymbol\upbeta=0$. For ease of notation, we define the first and second derivatives of $F_{n,i}(\cdot)$ as follows respectively
\begin{equation}\label{d1} 
F_{n,i}^{(1)}\left( u \right) = \frac{{\partial {F_{n,i}}\left( u \right)}}{{\partial u}} = {\varphi _{{w_n}}}\!\!\left( {{\tau _{i -\! 1}} - u} \right) - {\varphi _{{w_n}}}\!\!\left( {{\tau _i} - u} \right),
\end{equation}
and
\begin{equation} \label{d2} 
\begin{split}
\!\!F_{n,i}^{(2)}\left( u \right) &= \frac{{\partial F_{n,i}^{\left( 1 \right)}\left( u \right)}}{{\partial u}}\\
&= \frac{{{\tau _{i -\! 1}}\! - \!u}}{{\sigma _n^2/2}}{\varphi _{{w_n}}}({{\tau _{i -\! 1}} \!-\! u})- \frac{{{\tau _i}\! - \!u}}{{\sigma _n^2/2}}{\varphi _{{w_n}}}({\tau_i}-u),
\end{split}
\end{equation}
where $\varphi_{w_n}$ represents the probability density function of $\Re({w_n})$ or $\Im({w_n})$. And we simplify the expressions as $F_{n,i}(0)=F_{n,i},F_{n,i}^{(1)}(0) = F_{n,i}^{(1)}$ and $F_{n,i}^{(2)}(0) = F_{n,i}^{(2)}$. Then we obtained the closed-form $\Lambda _R $ reported in (\ref{prof}), which is shown at bottom of this page (the proof is given in Appendix). Given $q=1$ and $\tau_1 = 0$, the formula can be further simplified, and the specific form of (\ref{prof}) is shown in \cite{radar3}.
\begin{figure*}[bp]
\hrulefill
\begin{equation}\label{prof}
\Lambda _R\! = \! \dfrac{{{{\left\{ {\displaystyle\sum\limits_{n = 1}^{{N_q}}\! {\left[ {\displaystyle\sum\limits_{i = 1}^{{2^q}} \!{\dfrac{{{g_n}{I_i}\left( {\Re \left( {{y_n}} \right)} \right)F_{n,i}^{\left( 1 \right)}}}{{{F_{n,i}}}}} } \right.} \left. { \!+\! \displaystyle\sum\limits_{i = 1}^{{2^q}} \!{\dfrac{{{h_n}{I_i}\left( {\Im \left( {{y_n}} \right)} \right)F_{n,i}^{\left( 1 \right)}}}{{{F_{n,i}}}}} } \right]} \right\}}^2}\! \!\!+\! {{\left\{ {\displaystyle\sum\limits_{n = 1}^{{N_q}}\! {\left[ {\displaystyle\sum\limits_{i = 1}^{{2^q}}\! {\dfrac{{{g_n}{I_i}\left( {\Im \left( {{y_n}} \right)} \right)F_{n,i}^{\left( 1 \right)}}}{{{F_{n,i}}}}} \!-\! \displaystyle\sum\limits_{i = 1}^{{2^q}}\! {\dfrac{{{h_n}{I_i}\left( {\Re \left( {{y_n}} \right)} \right)F_{n,i}^{\left( 1 \right)}}}{{{F_{n,i}}}}} } \right]} } \right\}}^2}}}{{\displaystyle\sum\limits_{n = 1}^{{N_q}} {\left\{ {\left[ {g_n^2 + h_n^2} \right]\displaystyle\sum\limits_{i = 1}^{{2^q}} {{\dfrac{{{{\left( {F_{n,i}^{\left( 1 \right)}} \right)}^2} - F_{n,i}^{\left( 2 \right)}{F_{n,i}}}}{{{F_{n,i}}}}}} } \right\}} }}
\end{equation}
\end{figure*}
\section{Performance Analysis and Quantizer Design}
In this section, we first state results for the asymptotic performances of the GLRT and Rao test. According to \cite{ss2}, the test statistic ${\Lambda _R}$ (as well as ${\Lambda _G}$) is asymptotically distributed as
\begin{equation}\label{distribution}
{\Lambda_R}\mathop \sim \limits^\alpha \left\{ \begin{array}{l}
\chi _2^2,\quad\quad\quad\quad {H_0},\\
\chi _2^{{\prime2}}\left( {{\lambda _F}} \right),\quad\;\,{H_1},
\end{array} \right.
\end{equation}
where $\alpha$ denotes an asymptotic PDF, $\chi _2^2$ denotes the chi-square distribution with 2 degrees of freedom, $\chi^{\prime2}_2\left(\cdot\right)$ denotes the non-central chi-squared distribution with 2 degrees of freedom. $\lambda_F$ is the non-centrality parameter, such that 
\begin{equation}\label{lamda}
\begin{aligned}
{\lambda _F}&\!=\!({\boldsymbol\upbeta}_1-{\boldsymbol\upbeta}_0)^{\top}{\bf{FI}}({\boldsymbol\upbeta}_0)({\boldsymbol\upbeta}_1-{\boldsymbol\upbeta}_0)\\ &\!=\!{{\left \| {\boldsymbol\upbeta}_1 \right \|}^2}{{\sum\limits_{n = 1}^{{N_q}}\!\! {\left\{\! {\left[ {g_n^2 + h_n^2} \right]\sum\limits_{i = 1}^{{2^q}} {\left[ {\frac{{{{\left( {F_{n,i}^{\left( 1 \right)}} \right)}^2}\!\! -\!\! F_{n,i}^{\left( 2 \right)}{F_{n,i}}}}{{{F_{n,i}}}}} \right]} } \!\right\}} }},
\end{aligned}
\end{equation}
where ${\boldsymbol\upbeta}_1$ denotes the truth value of ${\boldsymbol\upbeta}$ under $H_1$. According to the characteristics of noncentral chi-square distribution, the detection performance will get better with the increase of $\lambda _F$ when global threshold $\eta$ is given by the false-alarm probability. Then the quantization thresholds $\left\{\tau_k,k = 0,1,...,{2^q} \right\}$ would be the only controllable variables of (\ref{lamda}). Thus the asymptotic detection performance of the Rao test can be further improved by the optimization of quantization thresholds, which are obtained as ($\tau _0^* = - \infty$ and $\tau_{2^q}^* = + \infty$ always)
\begin{equation}\label{op}
\begin{array}{l}
\left\{ {\tau _k^*,k = 1,...,{2^q} - 1} \right\} = \arg \mathop {\max }\limits_{\tau _k}{\lambda _F}\\
\quad\quad{\text{s.t.}}\quad{\tau _1} < {\tau _2} < \cdots < {\tau _{{2^q} - 1}}
\end{array}.
\end{equation}
\quad Since the above optimization problem does not allow a closed-form analytic solution, the numerical algorithm would be a feasible method. In this paper, we obtain the approximate results by utilizing the PSOA, which does not rely on the concavity property \cite{m3}. The PSOA is a random search algorithm inspired by the foraging behavior of birds, which aims at optimizing a problem by iteratively improving a set of particles\cite{PSO1}. For our case, the $\lambda_F$ is chosen as the objective function, remaining optimization parameters and steps are the same as those in reference \cite{m1}.

Now, the asymptotic distributions of test statistic are available because the $\lambda _{F\max }$ is obtained by substituting $\left\{\tau_k^*,k = 0,1,...,{2^q} \right\}$ into (\ref{distribution}). we proceed by deriving the detection probability $P_D$ for a given false-alarm probability $P_{FA}$. From (\ref{distribution}), the detection probability $P_D$ for multi-bit detector is approximated as
\begin{equation}
{P_D} = 1 - {Q_{\chi _2^{\prime2}\left( {{\lambda _{F\max}}} \right)}}\left( {Q_{_{\chi _2^2}}^{ - 1}\left( {1 - {P_{FA}}} \right)} \right),
\end{equation}
where $Q_{\chi _2^{\prime2}\left( {{\lambda _{F\max}}} \right)}$ and $Q_{\chi _2^2}$ represent the noncentral chi-square and chi-square cumulative distribution function respectively \cite{ss2}. In our case, the quantization thresholds for each signal element are considered to be the same, which could be different for a better performance theoretically. Since the performance gain by quantizing each signal element differently is very limited in low SNR, we chose the simplified model for decrease of computation.
\begin{table}[]
\small
\centering
\caption{Quantizer design for $q$-bit detectors}
\renewcommand{\arraystretch}{1.2} 
\begin{tabular}{|b{1.5cm}<{\centering}|b{0.8cm}<{\centering}|b{1.5cm}<{\centering}|b{0.8cm}<{\centering}|} 
\hline
\multicolumn{4}{|c|}{Quantization thresholds obtained by the PSOA}\\
\multicolumn{4}{|c|}{(${\text{SNR}} = -14{\text{dB}}$, $\sigma_w^2=2$)} \\ \hline
\multicolumn{2}{|c|}{\multirow{2}{*}{\begin{tabular}[c]{@{}c@{}}$\tau_0^*=-\infty$\\$\tau_{2^q}^*=+\infty$\end{tabular}}} &
\multirow{7}{*}{\begin{tabular}[c]{@{}c@{}}\!$q=3$\\ \!\!$(\tau_1^*,\cdots,\tau_7^*)$\end{tabular}} &
\multirow{7}{*}{\begin{tabular}[c]{@{}c@{}}-1.630\\ -1.012\\-0.460\\ 0.067\\ 0.542\\ 1.067\\ 1.803\end{tabular}} \\
\multicolumn{2}{|c|}{} & & \\ \cline{1-2}
\multirow{2}{*}{\begin{tabular}[c]{@{}c@{}}$q=1$\\$(\tau_1^*)$ \end{tabular}} & \multirow{2}{*}{-0.003} & & \\
& & & \\ \cline{1-2}
\multirow{3}{*}{\begin{tabular}[c]{@{}c@{}}$q=2$\\ $(\tau_1^*,\tau_2^*,\tau_3^*)$\end{tabular}} &
\multirow{3}{*}{\begin{tabular}[c]{@{}c@{}}-0.978\\-0.008\\0.967\end{tabular}} &
&
\\
& & & \\
& & & \\ \hline
\end{tabular}
\end{table}
\section{Simulation Results}
In this section, we delve into examining the performances of the multi-bit detectors by utilizing numerical simulation. The classic $\infty$-bit (unquantized signal) GLRT detector is considered as the upper bound \cite{ss2}, whose corresponding test statistic is given by
\begin{equation}
{\Lambda _{G - \infty }} = \frac{{\hat{\boldsymbol\upbeta}}_\infty^H{\bf{z}}^H{\bf{z}}{\hat{\boldsymbol\upbeta}}_\infty}{{\sigma_n ^2}/2},
\end{equation}
where ${{\hat{\boldsymbol\upbeta}}_\infty}\!\!=\!\!{\left( {{{\bf{z}}^H}{\bf{z}}} \right)^{ - 1}}{{\bf{z}}^H}{\bf{x}}$ is the MLE of ${\boldsymbol\upbeta}$ based on the unquantized signal, the vectors $\bf{z}$ and $\bf{x}$ represent the vectorized forms of the matrices $\bf{Z}$ and $\bf{X}$, respectively. And the non-centrality parameter for $\infty$-bit (unquantified signal) GLRT detector is calculated as
\begin{equation}
{\lambda _{F - \infty } = \frac{{{\boldsymbol\upbeta}}_1^H{\bf{z}}^H{\bf{z}}{{\boldsymbol\upbeta}}_1}{{\sigma_n ^2}/2}}.
\end{equation}

In the simulation, we define the target signal-to-noise ratio as ${\text{SNR}} = 10{\log _{10}}({\left \| {\boldsymbol\upbeta} \right \|}^2/{\sigma _n^2})$, and adopt the orthogonal linear frequency modulation (LFM) signal as the transmitted waveform \cite{MIMO3,MIMO4,MIMO5}, which is expressed as 
\begin{equation}
{\bf{S}}\left( {p,l} \right) = \frac{\exp \left\{ {{{j2\pi p\left( {l - 1} \right)} \mathord{\left/
{\vphantom {{j2\pi p\left( {l - 1} \right)} L}} \right.
\kern-\nulldelimiterspace} L} + {{j\pi {{\left( {l - 1} \right)}^2}} \mathord{\left/
{\vphantom {{j\pi {{\left( {l - 1} \right)}^2}} L}} \right.
\kern-\nulldelimiterspace} L}} \right\}}{N_t},
\end{equation}
where $p = 1,...,{N_t}$ and $l = 1,...,L$. It should be noted that the signal element, instead of the signal matrix, is normalized for analyzing the impact of sample size on detection performance conveniently.

To show the quantization details, we first list the quantization thresholds for $q$-bit detectors $(q=1,q=2$ and $q=3)$ in TABEL $\text{\uppercase\expandafter{\romannumeral1}}$, which are obtained by the PSOA with $N_r\times L = 16 \times 8$. Significantly, the $\left\{\tau_k^*,k = 0,1,...,{2^q} \right\}$ are not designed by equal interval, and the central quantization thresholds for all multi-bit detectors are close to zero. By further comparing the differences between adjacent quantization thresholds, it can be seen that the optimal quantization thresholds obey a symmetric distribution, which is consistent with the results of real-valued case in \cite{m1,m2,m3}.
\begin{figure}
\centering
{\includegraphics[height = 6.5cm,width=8cm]
{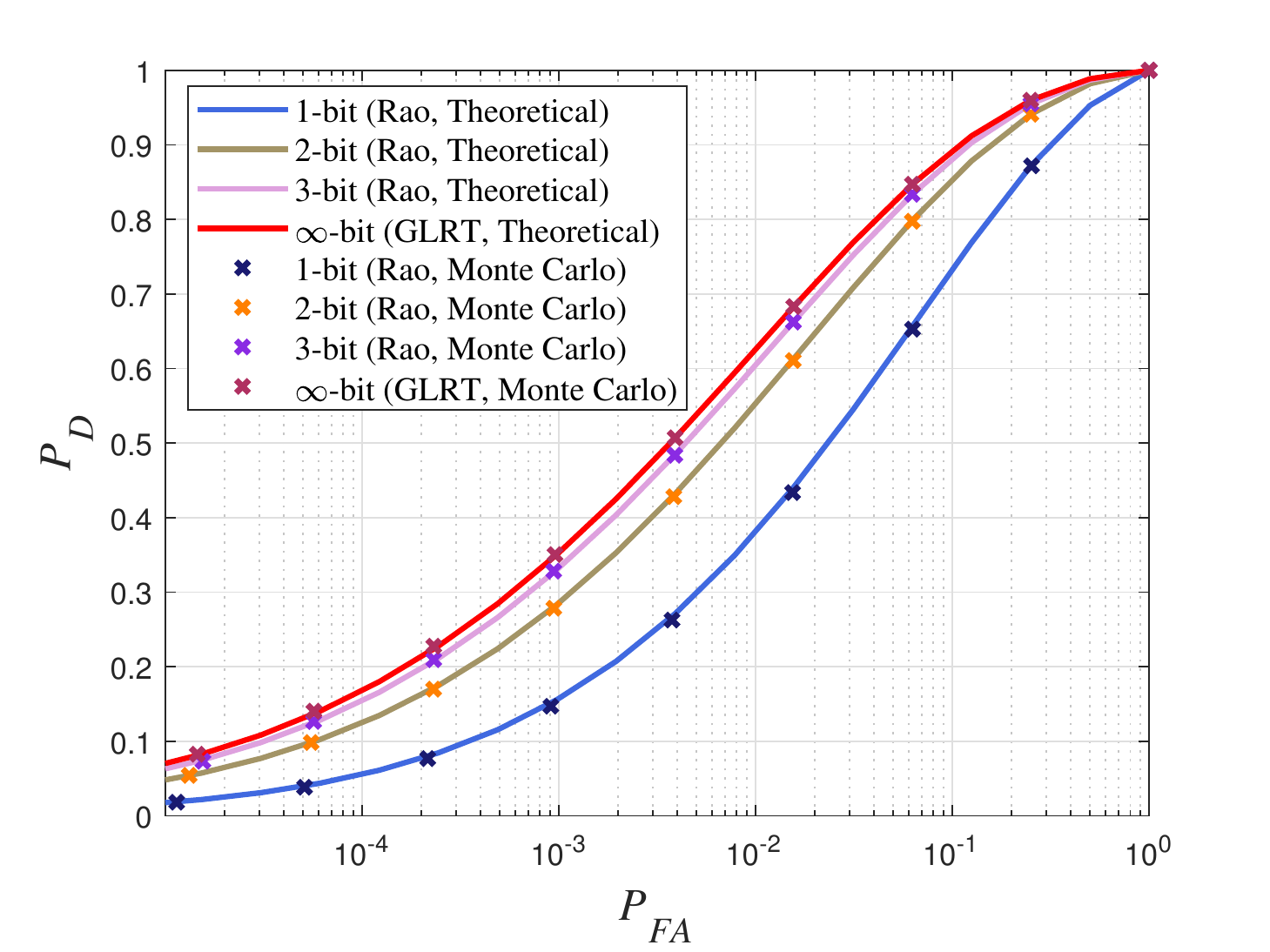}}
\caption{The receiver operating characteristic (ROC) curves for multi-bit detectors with ${\text{SNR}} = -14{\text{dB}}$, $N_r\times L = 16 \times 8$.}
\end{figure}
\begin{figure}
\centering
{\includegraphics[height = 6.5cm,width=8cm]
{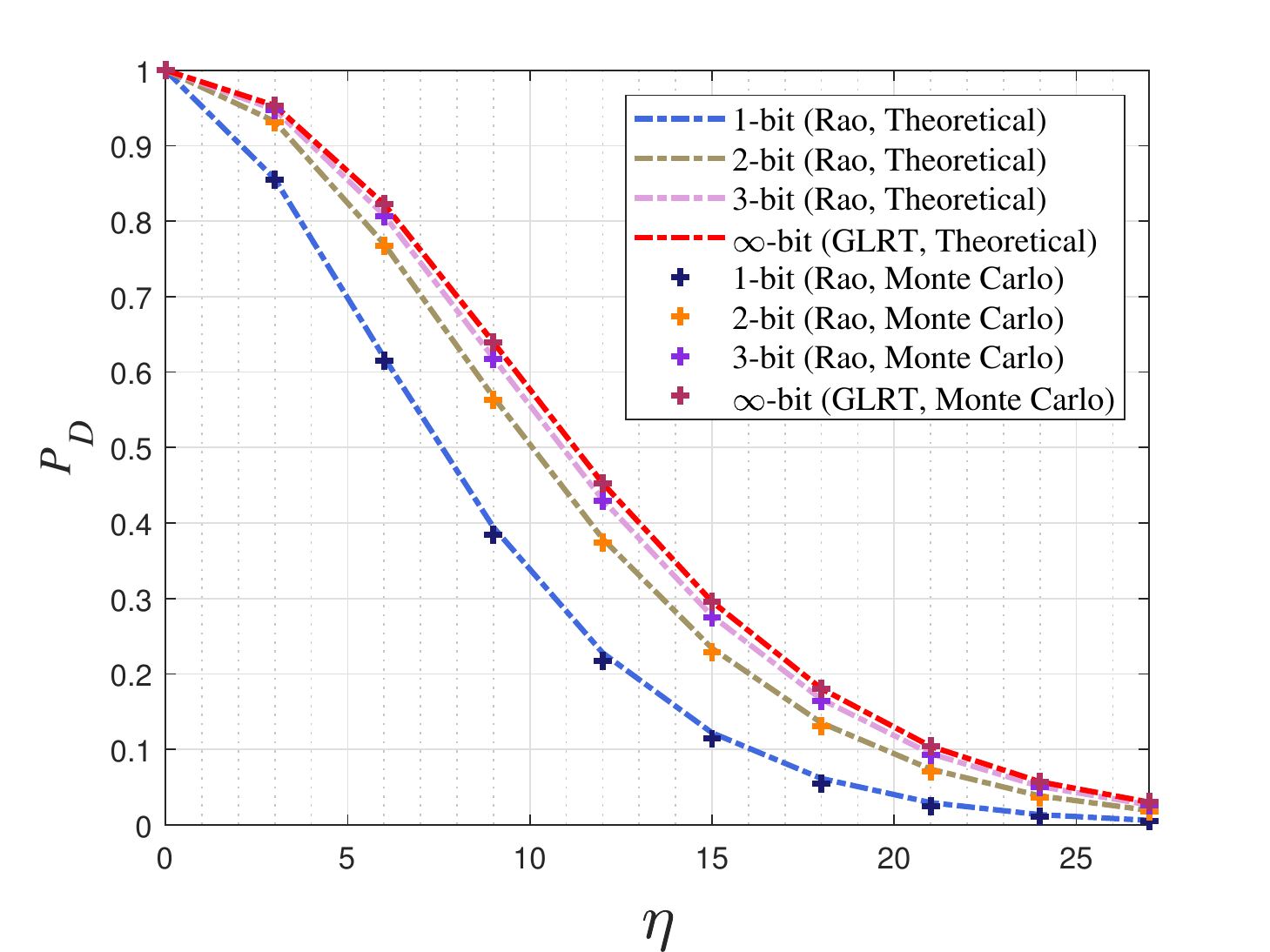}}
\caption{Detection probability $P_D$ versus global threshold $\eta$ for multi-bit detectors with ${\text{SNR}} = -14{\text{dB}}$, $N_r\times L = 16 \times 8$.} 
\end{figure}

Then, we apply the above quantization thresholds to corresponding quantizers, and simulate the detection performance based on the same sample data. In Fig.1, the ROC curves for multi-bit detectors are plotted where ${\text{SNR}} = -14{\text{dB}}$, $N_r\times L = 16 \times 8$. From this we can see, reserving the information of original signal in maximum, the $\infty$-bit detector provides the best detection performance. In contrast, the 1-bit detector produces the most performance degradation caused by the simplest quantization. And the performances of other multi-bit detectors are between those of $\infty$-bit and 1-bit detectors. Especially the performance cost of 3-bit detector is almost disregarded. On the other hand, it is observed that our theoretical analysis and the Monte Carlo results (with $10^6$ trials) for multi-bit detectors are quite consistent, which indicates that the experimental performance can be well generalized by theoretical asymptotic performance. Also, we plot the detection probability $P_D$ versus the threshold $\eta$ for multi-bit detectors with identical parameters. The results show a similar performance variation, which further proves the effectiveness of the proposed multi-bit quantizer.

\begin{figure}
\centering
{\includegraphics[height = 6.5cm,width=8cm]
{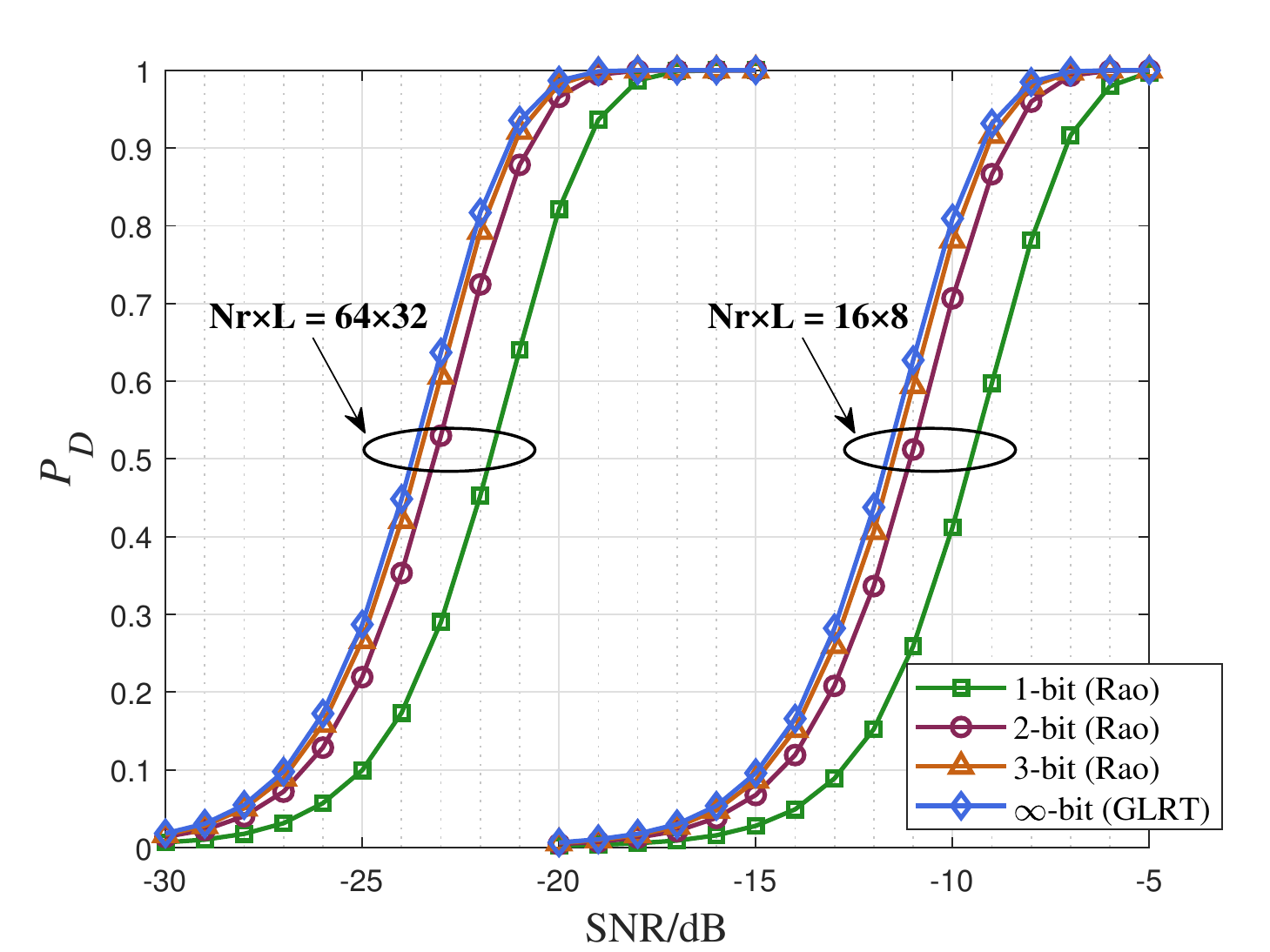}}
\caption {Detection probability (Monte Carlo) versus SNR for multi-bit detector with ${{P_{FA}}} = 10^{-4}$.} 
\end{figure}

As a complementary analysis, the impact of remaining parameters on performance is considered in our simulation. In Fig 5, we plot the detection probability $P_D$ versus the SNR for $P_{FA}=10^{-4}$, As shown in the figure, the performance of the 1-bit detector case is 2 dB weaker than the case of $\infty$-bit detector, which is in accordance with the results in \cite{radar3}. The results demonstrate that the increase of the bit depth leads a gain in the detection performance as same as the conclusion about Fig 1. Additionally, by comparing the results with different sample sizes, it can be seen almost consistent performance trends, which means the improvement of detection performance is equal to the increase of sample size for multi-bit detector in low SNR.
\section{Conclusion}
To reduce complexity of data processing in large-scale radar systems, we propose a detector on Rao test with multi-bit quantization, which applies to a weak target detection with unknown parameters. The test statistic has a closed form, whose theoretical asymptotic distribution is provided. Also, we obtained the optimized thresholds by maximizing the noncentral parameter quantization. Simulation results are consistent with the theoretical analyses and show that the $\infty$-bit and 1-bit detectors expose the best and worst performances respectively due to their extreme quantization methods. The increase of quantization bits leads a performance improvement, which proves the effectiveness of the proposed multi-bit quantizer.
\begin{appendix}
\section{Proof of }
Firstly, we express $\ln P({\left.{\bf{Y}}\right|{H_1};{\boldsymbol{\upbeta}}})$ as $\ell(\bf{Y};{\boldsymbol{\upbeta}})$. Based on the definition of (\ref{d1}) and (\ref{d2}), the partial derivatives of the likelihood function with respect to $\boldsymbol\upbeta$ are given by respectively
\begin{equation}\label{proof1}
\begin{aligned}
\!\!\!\!{\frac{{\partial \ell(\bf{Y};{\boldsymbol{\upbeta}})}}{{\partial {\beta _R}}}} \Bigg|_{{\boldsymbol\upbeta} = {\boldsymbol\upbeta}_0}\!\! = \! \sum\limits_{n = 1}^{N_q\!\times\!L}\Bigg[\!&\sum\limits_{i = 1}^{2^q}\frac{{g_n{I_i}(\Re ( y_n))F_{n,i}^{(1)}}}{F_{n,i}}\\
&\quad\quad + \sum\limits_{i = 1}^{{2^q}} {\frac{{{h_n}{I_i}\left( {\Im \left( {{y_n}} \right)} \right)F_{n,i}^{\left( 1 \right)}}}{{{F_{n,i}}}}}\!\Bigg]
\end{aligned}
\end{equation}
and
\begin{equation}\label{proof2}
\begin{aligned}
\!\!\!\!\!\!{\frac{{\partial \ell(\bf{Y};{\boldsymbol{\upbeta}})}}{{\partial {\beta _I}}}} \Bigg|_{{\boldsymbol\upbeta} = {\boldsymbol\upbeta}_0}\!\! = \! \sum\limits_{n = 1}^{N_q\!\times\! L}\Bigg[\!&\sum\limits_{i = 1}^{2^q}\frac{-{h_n{I_i}(\Re ( y_n))F_{n,i}^{(1)}}}{F_{n,i}}\\
&\quad\quad + \sum\limits_{i = 1}^{{2^q}} {\frac{{{g_n}{I_i}\left( {\Im \left( {{y_n}} \right)} \right)F_{n,i}^{\left( 1 \right)}}}{{{F_{n,i}}}}}\!\Bigg]\!.
\end{aligned}
\end{equation}
\quad Then we compute $\bf{FI}({\boldsymbol\upbeta}_0)$ element by element, as follows
\begin{equation}\label{proof3}
\begin{aligned}
{\bf{FI}}_{1,1}({\boldsymbol\upbeta}_0)&=\mathbb{E}\left[-\frac{{\partial^2 \ell(\bf{Y};{\boldsymbol{\upbeta}})}}{{\partial {\beta _R}^2}}\right]\Bigg|_{{\boldsymbol\upbeta}={\boldsymbol\upbeta}_0}\\
&=\mathbb{E}\left[\sum\limits_{n = 1}^{N_q\!\times\! L}\!\left\{\!\sum\limits_{i = 1}^{2^q}\frac {g_n^2I_i(\Re(y_n))\left[\!\left(\!F_{n,i}^{(1)}\!\right)^2 \!\!\!-\!{F_{n,i}^{(2)}F_{n,i}}\!\right]}{\left(F_{n,i}\right)^2} \right.\right.\\ 
&\qquad\quad+\!\left.\left.\sum\limits_{i = 1}^{2^q}\frac {h_n^2I_i(\Im(y_n))\left[\!\left(\!F_{n,i}^{(1)}\!\right)^2\!\!\!-\!{F_{n,i}^{(2)}F_{n,i}}\!\right]}{\left(F_{n,i}\right)^2}\!\right\}\right]\\
&={{\sum\limits_{n = 1}^{{N_q}} {\left\{ {\left[ {g_n^2 + h_n^2} \right]\sum\limits_{i = 1}^{{2^q}} {{\dfrac{{{{\left(\! {F_{n,i}^{\left( 1 \right)}} \!\right)}^2}\!\! - \!F_{n,i}^{\left( 2 \right)}{F_{n,i}}}}{{{F_{n,i}}}}}} } \right\}} }}\\
&=\mathbb{E}\left[-\frac{{\partial^2 \ell(\bf{Y};{\boldsymbol{\upbeta}})}}{{\partial {\beta _I}^2}}\right]\Bigg|_{{\boldsymbol\upbeta}={\boldsymbol\upbeta}_0}={\bf{FI}}_{2,2}({\boldsymbol\upbeta}_0)
\end{aligned}
\end{equation}
and
\begin{equation}\label{proof4}
\begin{aligned}
{\bf{FI}}_{1,2}({\boldsymbol\upbeta}_0)&=\mathbb{E}\left[-\frac{{\partial^2 \ell(\bf{Y};{\boldsymbol{\upbeta}})}}{{\partial {\beta _R}\partial {\beta _I}}}\right]\Bigg|_{{\boldsymbol\upbeta}={\boldsymbol\upbeta}_0}\\
&=\mathbb{E}\left[\sum\limits_{n = 1}^{N_q\!\times\! L}\!\left\{\!\sum\limits_{i = 1}^{2^q}\frac {-I_i(\Re(y_n))\left[\!\left(\!F_{n,i}^{(1)}\!\right)^2 \!\!\!-\!{F_{n,i}^{(2)}F_{n,i}}\!\right]}{\left(\!F_{n,i}\!\right)^2} \right.\right.\\ 
&\quad\;+\!\left.\left.\sum\limits_{i = 1}^{2^q}\frac {I_i(\Im(y_n))\left[\!\left(\!F_{n,i}^{(1)}\!\right)^2\!\!\!-\!{F_{n,i}^{(2)}F_{n,i}}\!\right]}{\left(\!F_{n,i}\!\right)^2}\!\right\}g_nh_n\right]\\
&=0={\bf{FI}}_{2,1}({\boldsymbol\upbeta}_0).
\end{aligned}
\end{equation}
\quad From the results of $\bf{FI}({\boldsymbol\upbeta}_0)$, (\ref{1}) have a more
simplified expression as
\begin{equation}\label{proof5}
{\Lambda_R} = \frac{\left[{\dfrac{{\partial \ell(\bf{Y};{\boldsymbol{\upbeta}})}}{{\partial {\beta _R}}}} \Bigg|_{{\boldsymbol\upbeta} = {\boldsymbol\upbeta}_0}\right]^2 + \left[{\dfrac{{\partial \ell(\bf{Y};{\boldsymbol{\upbeta}})}}{{\partial {\beta _I}}}} \Bigg|_{{\boldsymbol\upbeta} = {\boldsymbol\upbeta}_0}\right]^2}{{\bf{FI}}_{1,1}({\boldsymbol\upbeta}_0)}.
\end{equation}

Finally, the desired result in (\ref{prof}) is obtained by substituting (\ref{proof1}) (\ref{proof2}) and (\ref{proof3}) into (\ref{proof5}).
\end{appendix}
\balance

\vspace{12pt}
\color{red}
\end{document}